\documentclass[12pt, titlepage]{article}
\usepackage{a4wide, amssymb, latexsym, epsfig}

\newcommand{\ga}{\alpha}

\newcommand{\eps}{\epsilon}

\newcommand{\gl}{\lambda}

\newcommand{\gD}{\Delta}

\renewcommand{\hbar}{\overline{h}}

\newcommand{\Ghat}{\hat{G}}
\newcommand{\Mhat}{\hat{M}}

\newcommand{\tauhat}{\hat{\tau}}

\newcommand{\cM}{\mathcal M}

\newcommand{\bR}{\mathbb R}

\newcommand{\bZ}{\mathbb Z}

\newcommand{\be}{\begin{equation}}
\newcommand{\bea}{\begin{eqnarray}}

\renewcommand{\d}{\partial}
\newcommand{\ee}{\end{equation}}
\newcommand{\eea}{\end{eqnarray}}

\newcommand{\ret}{\nonumber \\}

\begin{document}

\begin{titlepage}
 \begin{flushright}
  ITFA-2003-1 \\
  SPIN-2002/48 \\
  PUPT-2072 \\[4cm]
 \end{flushright}
 \begin{center}
  {\bf \Large String networks and supersheets} \\[1cm]
  Erik Verlinde\footnote{Physics Department, Princeton University, 
  Princeton, NJ 08544, USA} and Marcel Vonk\footnote{Institute for Theoretical
  Physics, University of Amsterdam, 1018 XE Amsterdam, The 
  Netherlands}$^,$\footnote{Spinoza Institute, Utrecht University, Leuvenlaan 4,
  3584 CE Utrecht, The Netherlands} \\[4cm]
  {\bf Abstract} \\[1ex]
 \end{center}
 As was shown by Aharony, Hanany and Kol and independently by Sen, type IIB 
 string theory admits configurations where strings of different charges $(p_i, 
 q_i)$ form so-called string networks. We argue that these networks blow up into
 ``supersheets'': supersymmetric spinning cylindrical D3-branes carrying 
 electric and magnetic fluxes. These supersheets are three-dimensional 
 generalizations of the supertubes that were constructed by Mateos and Townsend.
 We calculate the mass of both systems for arbitrary values of the parameters 
 and find exact agreement.
\end{titlepage}

\section{Introduction}
\setcounter{footnote}{0}
In \cite{Mateos:2001st}, Mateos and Townsend showed that D2-branes in type IIA
string theory admit stable supersymmetric configurations of cylindrical 
topology, the so-called supertubes. These supertubes are supported against 
collapse by an angular momentum, which is the result of a combined electric and
magnetic field on the world volume. It was noted that a supertube can
be obtained by starting from a sequence of D0-branes placed on a long string, 
like beads on a necklace. This structure then starts spinning around the 
longitudinal axis and blows up into a supertube.

Several generalizations of the supertube have been studied; see e.\ g.\ [2-10].
In this paper, we consider what happens in a T-dual type IIB picture, where the
T-duality acts along a perpendicular direction. (The case of a T-duality along
the supertube is also very interesting; see \cite{Cho:2001sd, Mateos:2002sc}). 
Here, one would expect the D0-branes to turn into D-strings which are 
intertwined with fundamental strings. In fact, in type IIB string theory more 
general $(p,q)$-strings exist, so there are many other possibilities as 
well.\footnote{The relation between $(p,q)$-strings and D3-supertubes was also 
discussed in \cite{Tamaryan:2002td}} These ``string networks'' were first 
described by Aharony, Hanany and Kol in \cite{Aharony:1998wb} and independently
by Sen in \cite{Sen:1998sn}.

In this paper, we claim that these string networks indeed blow up into
D3-branes that are a direct generalization of supertubes.\footnote{In 
\cite{Kumar:2001bs}, a bound state of a D3-brane with a string network was 
considered. However, here we claim that the string network {\em itself} blows 
up into a D3-brane.} It is as if the string network ``fills up'' and obtains a 
small  but finite thickness. Because of this sheet-like structure, we use the 
term ``supersheets'' to describe these D3-branes. We claim that the precise 
structure of the string network is not important, but that only the effective 
electric and magnetic fields it produces determine the type of resulting 
supersheet. To support these claims, we calculate the mass of both objects for 
arbitrary parameters, and obtain exact agreement.

The outline of this paper is as follows. In section 2, we give a very brief
review of string networks. Our main goals here are to fix notation and to give
the mass formula that was found by Sen; for a more detailed introduction the
reader is referred to the original papers \cite{Aharony:1998wb, Sen:1998sn}. 
In section 3 we calculate the supersheet Hamiltonian in a way that is analogous 
to the supertube calculation by Mateos and Townsend. Section 4 contains the 
proof that the mass of such a supersheet is exactly equal to the mass of the 
corresponding string network. In section 5, we argue that there is a 
many-to-one mapping back from string networks to supersheets. Finally, section 
6 contains our conclusions.

\section{Brief review of string networks}
\label{sec:stringnetworks}
It is well known that type IIB string theory is invariant under an $SL(2, \bZ)$
duality group. One of the consequences of this fact is that the theory does not
only have fundamental strings (F-strings) and D-strings, but that it also 
contains $(p,q)$-strings which have fundamental string charge $p$ and D-string 
charge $q$ whenever $p$ and $q$ are relatively prime \cite{Witten:1996bs}. It 
was shown by Aharony, Hanany and Kol \cite{Aharony:1998wb} and independently by 
Sen \cite{Sen:1998sn} that these strings can form two-dimensional networks, 
where at each three-point vertex the charge of the strings is conserved, i.\ 
e.\ a $(p_1,q_1)$-string can split into a $(p_2, q_2)$-string and a $(p_3, 
q_3)$-string if $p_1 = p_2 + p_3$ and $q_1 = q_2 + q_3$. Using the fact that 
the tension of a string is known to be \cite{Schwarz:1995ms}
\be
 T_{p,q} = \frac{1}{\gl_2} | p + q \gl|,
\ee
where $\gl = \gl_1 + i \gl_2$ is the complex coupling constant of type IIB
string theory, it was shown that such a network is in equilibrium (i.\
e.\ the force on each vertex is zero) if every $(p,q)$-string has the direction
$p + q \gl$ in complex coordinates. 
\begin{figure}[h]
 \begin{center}
  \epsfysize=4 cm
  \epsfbox{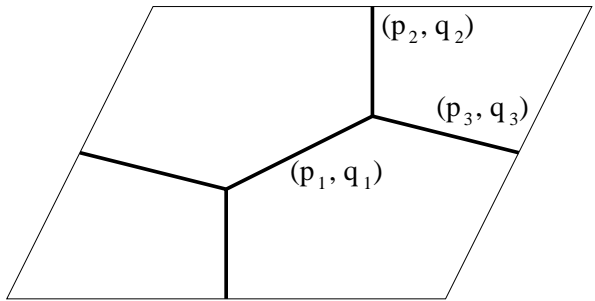}
 \end{center}
 \caption{A string network consisting of three strings on a torus.}
 \label{fig:network}
\end{figure}
The simplest case of such a network is one consisting of only three strings on 
a torus of modular parameter $\tau$, as in figure \ref{fig:network}. Sen showed 
that the mass of such a network can be calculated, and it is given by
\be
 M^2 = A \left(
  \begin{array}{c}
   p_1 \\  q_1 \\ p_2 \\ q_2
  \end{array}
 \right)^T (\pm L + \Mhat)
 \left(
  \begin{array}{c}
   p_1 \\  q_1 \\ p_2 \\ q_2
  \end{array}
 \right),
 \label{eq:masssn}
\ee
where $A$ is the area of the torus; $(p_i, q_i)$ are the charges of two of the
three strings (the charges of the third one can of course be expressed in terms 
of these), and $L$ and $\Mhat$ are the following matrices:
\be
  L = 
 \left(
  \begin{array}{rrrr}
   0 & 0  & 0  & 1 \\
   0 & 0  & -1 & 0 \\
   0 & -1 & 0  & 0 \\
   1 & 0  & 0  & 0 \\
  \end{array}
 \right), \qquad
 \Mhat = \cM_{\tau} \otimes \cM_\gl,
\ee
with
\be
 \cM_\gl = \frac{1}{\gl_2} \left( \begin{array}{cc} 1 & \gl_1 \\ \gl_1 &
 |\gl|^2 \end{array} \right),
 \label{eq:SL2Zinv}
\ee
and similarly for $\cM_\tau$. Here, we adopted the convention that the first 
factor in a tensor product mixes the first two components of a four-vector with
the last two components, and the second factor mixes the first two components 
(and similarly the last two) among each other. Matrices of the type
(\ref{eq:SL2Zinv}) are $SL(2, \bZ)$-covariant, and give invariant quantities
when contracted with an $SL(2, \bZ)$-vector on both sides. Finally, the sign 
in front of $L$ in (\ref{eq:masssn}) is chosen such that the contribution of 
this term is positive.

\section{The supersheet Hamiltonian}
In this section we will derive the Hamiltonian for the three-dimensional 
supersheet. This calculation is analogous to the calculation for supertubes 
in \cite{Mateos:2001st}. More about the Born-Infeld Hamiltonian and its
dualities can be found in \cite{Hofman:1998ud}.

\subsection{The system}
\label{sec:system}
We consider type IIB string theory on a space-time of the form
\be
 M_{10} = \bR_T \times T^2 \times \bR^2 \times M_5.
\ee 
$\bR_T$ is the time-direction, parameterized by a parameter $T$. The torus 
$T^2$ is flat and has modular parameter $\tau$. To parameterize it, we will 
use coordinates $X, Y$ along perpendicular directions. The coordinate $X$ will 
be periodic with period $L_x$; the coordinate $Y$ will range from $0$ to $L_y$. 
This gives a cylinder, whose edges at $Y=0, L_y$ are identified with the 
appropriate twist to obtain the modular parameter $\tau$. $\bR^2$ is also
equipped with a flat metric, and polar coordinates $\Phi, R$. 
Finally, $M_5$ is an arbitrary 5-dimensional manifold which will play no further
role in the discussion below. 

Putting everything together, and adopting a ``mostly plus'' sign convention, 
the metric of our space-time in these coordinates is
\be
 ds^2 (M_{10}) = -dT^2 + dX^2 + dY^2 + dR^2 + R^2 d \Phi^2 + ds^2(M_5).
\ee 
We now wrap a D3-brane around the $T^2$, and give the remaining direction the
shape of a circle of fixed radius\footnote{To simplify the notation we 
will denote both the radial coordinate and the fixed radius at which the 
supersheet sits by the symbol $R$.} $R$ around the origin in the $\bR^2$-plane.
The world-volume coordinates of the three-brane are $(t, x, y, \phi)$, where we
take each of the last three coordinates to be periodic with periods $L_x, L_y$ 
and $2 \pi$ respectively. We fix the world-volume reparametrization invariance
by gauging 
\be
 \begin{array}{ccl}
   T & = & t      \\
   X & = & x + cy \\
   Y & = & y      \\
  \Phi & = & \phi
 \end{array}
\ee
From this embedding we see that the modular parameter of the space-time torus 
is related to the constants $c, L_x$ and $L_y$ by
\be
 \tau = (c + i) \frac{L_y}{L_x}.
\ee
A brief comment on our notation: world-volume quantities will have indices 
$i, j, k, \ldots$, where these indices range over the set $\{ t, x, y, \phi \}$.
In particular, $x^t = t, x^x = x,$ etc. If we do not consider the time
component, we will use indices $a, b, c, \ldots$ Space-time quantities are 
denoted by indices $\mu, \nu, \rho, \ldots$ Again, $X^T = T, X^X = X,$ etc.

\subsection{The Hamiltonian}
The world-volume theory on the D3-brane is given by the Born-Infeld Lagrangean
\be
 L = - \frac{1}{2 \pi g_s} \sqrt{- \det(G + F)}
 \label{eq:BILagrangean}
\ee
where we normalized the fields in such a way that there are no unnecessary
factors inside the square root, and the overall factor is chosen such that the
expanded Lagrangean has a factor of $1 / 4 \pi$ in front of the $F^2$-term, which
is the convenient normalization for self-duality (cf. \cite{Verlinde:1995ga}). 
Note that we also set $\ga'=1$, so all lengths in the following will be 
dimensionless quantities measured in terms of the string length. In this 
Lagrangean, $G_{ij}$ is the induced metric on the world-sheet:
\be
 G_{ij} \equiv \frac{\d X^\mu}{\d x^i} \frac{\d X^\nu}{\d x^j} G_{\mu \nu} = 
 \left(
  \begin{array}{cccc}
   -1 & 0 & 0       & 0   \\
    0 & 1 & c       & 0   \\
    0 & c & 1 + c^2 & 0   \\
    0 & 0 & 0       & R^2
  \end{array}
 \right)
 \label{eq:metric}
\ee
and $F_{ij}$ is the field strength of the world-sheet  $U(1)$ gauge-field 
$A_i$:
\be
 F_{ij} = 
 \left(
  \begin{array}{cccc}
    0        & -E_x      & -E_y        & -E_{\phi} \\
    E_x      &  0        &  R B^{\phi} & -R B^y    \\
    E_y      & -R B^\phi &  0          &  R B^x    \\
    E_{\phi} &  R B^y    & -R B^x      &  0
  \end{array}
 \right).
 \label{eq:fieldstrength}
\ee
The factors of $R$ in this expression are factors of $\sqrt{\det G}$ appearing
in the $\eps$-tensors that are used to define the magnetic field in terms of the
field strength.

Inserting these explicit matrices in (\ref{eq:BILagrangean}), we find after a
straightforward but slightly tedious calculation that
\be
 L = - \frac{1}{2 \pi g_s} \sqrt{1 + |B|^2 - |E|^2 - (B \cdot E)^2},
\ee
where the inner products are with respect to the space-like metric $G_{ab}$ and
its inverse $G^{ab}$. From this expression we can calculate the electric 
displacement
\be
 D^a \equiv 2 \pi \frac{\d L}{\d E_a} = - \frac{1}{g_s} \frac{G^{ab}E_b + B^a 
 (B \cdot E)}{\sqrt{1 + |B|^2 + |E|^2 + (B \cdot E)^2}},
\ee
where we introduced a factor of $2 \pi$ for future convenience, and the 
Hamiltonian
\be
 \label{eq:HinEandB}
 H \equiv \frac{D \cdot E}{2 \pi} - L = - \frac{1}{2 \pi g_s} \frac{1 + |B|^2} 
 {\sqrt{1 + |B|^2 + |E|^2 + (B \cdot E)^2}}.
\ee
Of course, this Hamiltonian should be rewritten in terms of $B$ and $D$. The
easiest way to do this is by considering the square of the Hamiltonian, and 
doing this it is not too difficult to see that
\be
 \label{eq:Hstringframe}
 H^2 = \frac{1}{(2 \pi)^2} \left\{ \frac{1}{g_s^2} + \frac{1}{g_s^2} |B|^2 + 
 |D|^2 + |D \times B|^2 \right\}.
\ee

\subsection{Einstein frame and axion field}
We want to modify the result (\ref{eq:Hstringframe}) in two ways. First of all,
we have worked in the string frame so far, to avoid cluttering our notation with
factors of the string coupling constant $g_s$. However, we would like to write
our final result in the Einstein frame. This means we redefine $G_{ij} \to
G_{ij} / \sqrt{g_s}$ to get rid of the coupling constant in front of the
$\sqrt{\det G}$-term in the action. From (\ref{eq:BILagrangean}) we see that 
this implies that our resulting Hamiltonian should be multiplied by a factor of
$g_s$, while all the gauge fields obtain a factor of $1/\sqrt{g_s}$. Note that
since $D^a$ is a derivative with respect to $E_a$, $D^a$ obtains a factor of 
$\sqrt{g_s}$.

Incorporating these redefinitions, we find that in the Einstein frame
\be
 \label{eq:Heinsteinframe}
 H^2 = \frac{1}{(2 \pi)^2} \left\{ 1 + \frac{1}{g_s} |B|^2 + g_s |D|^2 + |D 
 \times B|^2 \right\}.
\ee
Secondly, since we want to work with an arbitrary complex coupling constant, we
want to add an axion term
\be
 - \frac{\eta}{16 \pi} \eps^{ijkl} F_{ij} F_{kl} = - \frac{\eta}{2 \pi} E 
 \cdot B
\ee
to the Lagrangean (\ref{eq:BILagrangean}), where $\eta$ is the axion field. 
Since this term is linear in $E$, the functional form of the Hamiltonian in 
terms of $E$ and $B$ does not change. However, the expression for the electric 
displacement now changes by an amount $- \eta B^a$, so we expect the 
Hamiltonian in terms of $D$ and $B$ to be different from the case without axion.
In fact, we can simply compensate for this by replacing $D^a \to D^a + \eta B$
in the expression for the Hamiltonian in terms of $B$ and $D$, and we find
\be
 H^2 = \frac{1}{(2 \pi)^2} \left\{ 1 + \frac{1}{g_s} |B|^2 + g_s | D + \eta
 B|^2 + |D \times B|^2 \right\}.
\ee
We can now introduce the complex coupling constant
\be
 \gl = \gl_1 + i \gl_2 \equiv \eta + \frac{i}{g_s},
\ee
to rewrite this expression in the more elegant form
\be
 H^2 = \frac{1}{(2 \pi)^2} \left\{ 1 + \frac{1}{\gl_2} (D, \, B)
 \left(
  \begin{array}{cc}
   1 & \gl_1 \\
   \gl_1 & |\gl|^2
  \end{array}
 \right)
 \left(
  \begin{array}{c}
   D \\ B
  \end{array}
 \right)
 + |D \times B|^2 \right\}.
 \label{eq:finalhamiltonian}
\ee
This expression -- note the appearance of the matrix (\ref{eq:SL2Zinv}) -- is 
manifestly $SL(2, \bZ)$-invariant. Also note that this notation is somewhat
symbolic, in the sense that the matrix products also involve inner products
between the vectors.

\section{Connection to the string network}
So far, everything we did was completely general. Now we want to turn to the 
setup described in section \ref{sec:system}, to relate the supersheet to the
string network.

\subsection{Mass of the supersheet}
To this end, we take the $B$- and $D$-fields to be directed along the torus 
$T^2$. Since the torus is compact, these fields are quantized:
\bea
 && D^x = \frac{n^x}{R L_y} \qquad D^y = \frac{n^y}{R L_x} \ret
 && B^x = \frac{m^x}{R L_y} \qquad B^y = \frac{m^y}{R L_x},
\eea
for arbitrary integers $n^x, n^y, m^x$ and $m^y$. The total mass of the 
D-brane is now the Hamiltonian multiplied by the volume $V = 2 \pi R L_x L_y$:
\bea
 M^2 
 & = & V^2 H^2 \ret
 & = & R^2 L_x^2 L_y^2 + \frac{(n^x m^y - n^y m^x)^2}{R^2} + 
 \ret
 && \frac{1}{\gl_2} \left(
  \begin{array}{c}
   n^x L_x \\ n^y L_y \\ m^x L_x \\ m^y L_y
  \end{array}
 \right)^T
 \left(
  \begin{array}{cc}
   \Ghat       & \gl_1 \Ghat   \\
   \gl_1 \Ghat & |\gl^2| \Ghat
  \end{array}
 \right)
 \left(
  \begin{array}{c}
   n^x L_x \\ n^y L_y \\ m^x L_x \\ m^y L_y
  \end{array}
 \right),
 \label{eq:masssquaredst}
\eea
where $\Ghat$ is the metric restricted to the $x, y$-components:
\be
 \Ghat = \left(
  \begin{array}{cc}
   1 & c         \\
   c & 1 + c^2 \\
  \end{array}
 \right).
\ee
The idea of Mateos and Townsend \cite{Mateos:2001st} is that the supersheet 
will adjust its radius $R$ so that its energy becomes minimal. The radius for 
which this happens is
\be
 R^2 = \frac{\left| n^x m^y - n^y m^x \right|}{L_x L_y},
\ee
from which we can see that just like in the supertube case, the D3-brane does
not collapse due to its own tension, but it is supported by the angular momentum
coming from the electromagnetic field. We can insert this value for the radius
in (\ref{eq:masssquaredst}) to find
\bea
 M^2 = 2 L_x L_y |n^x m^y - n^y m^x| + \frac{1}{\gl_2} \left(
  \begin{array}{c}
   n^x L_x \\ n^y L_y \\ m^x L_x \\ m^y L_y
  \end{array}
 \right)^T
 \left(
  \begin{array}{cc}
   \Ghat       & \gl_1 \Ghat   \\
   \gl_1 \Ghat & |\gl^2| \Ghat
  \end{array}
 \right)
 \left(
  \begin{array}{c}
   n^x L_x \\ n^y L_y \\ m^x L_x \\ m^y L_y
  \end{array}
 \right).
 \label{eq:masssquaredstfinal}
\eea
This is our final expression for the mass of the supersheet, which we will 
compare to the mass of the corresponding string network in the next subsection.

\subsection{Mass of the string network}
\begin{figure}
 \begin{center}
  \epsfysize=4 cm
  \epsfbox{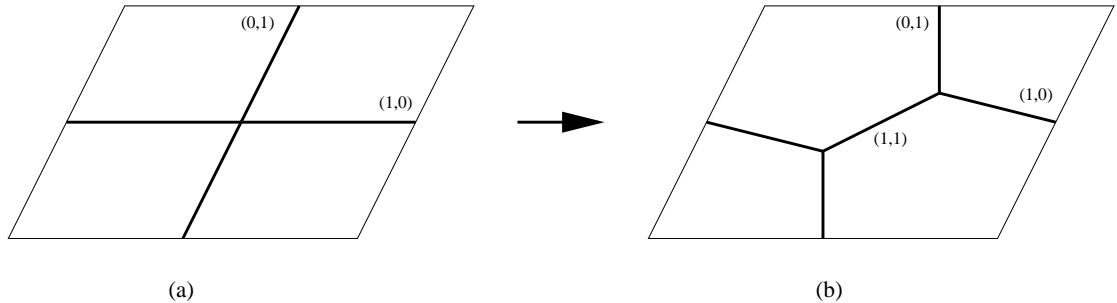}
 \end{center}
 \caption{The string network grows an extra string to achieve its preferred
 directions}
 \label{fig:grownstring}
\end{figure}

Our claim is that the supersheet corresponds to a ``blown up'' string network 
on the $(X,Y)$-torus. To establish the exact relation, note that a unit of
$D$-flux should correspond a fundamental string, and a unit of $B$-flux to a 
D-string. Hence we expect to find $(1,0)$-strings with winding numbers 
$(n^x, n^y)$ and $(0,1)$-strings with winding numbers $(m^x,
m^y)$\footnote{Strictly speaking, this is only true if $n^x$ and $n^y$ have no
common divisor, and similarly for $m^x$ and $m^y$. If for example $n^x$ and 
$n^y$ have a common divisor $d$, the correct description is in terms of $d$ 
fundamental strings of winding numbers $(n^x/d, n^y/d)$.}. In figure
\ref{fig:grownstring}a, we have drawn the simplest case of this, with winding 
numbers $(1,0)$ and $(0,1)$.
Of course, the directions of the strings in this network will in general not 
be the preferred directions $1$ and $\gl$ that were described in section 
\ref{sec:stringnetworks}. To achieve these preferred directions, the string 
network will ``grow a $(1,1)$-string'' as in figure \ref{fig:grownstring}b. 
However, since the mass formula (\ref{eq:masssn}) for the string network only 
depends on the total area of the torus and the quantum numbers of two of the 
three strings, we do not need to calculate the exact lengths of the resulting 
strings. Moreover, a counting of degrees of freedom shows that growing one extra
string is enough for the string network to achieve its preferred directions.

{\em The case with determinant 1}

For the moment we will assume that
\be
 \left| 
  \begin{array}{ll}
   n^x & n^y \\
   m^x & m^y
  \end{array}
 \right|
 = |n^x m^y - n^y m^x| = 1
 \label{eq:determinant}
\ee
so that the F- and D-strings (before the extra string grows) span a torus with 
an area equal to the area of the original torus. Hence by an $SL(2, 
\bZ)$-transformation we can view this network as a fundamental string of 
winding number (1,0) and a D-string with winding number (0,1) on a torus with 
sides $n^x L_x + n^y \tau L_x$ and $m^x L_x + m^y \tau L_x$. This new torus 
has modular parameter
\be
 \tauhat =
 \left(
  \begin{array}{rr}
   m^y & m^x \\
   n^y & n^x
  \end{array}
 \right)
 \cdot \tau, \qquad \mbox{or} \qquad
 \tau =
 \left(
  \begin{array}{rr}
   n^x & -m^x \\
   -n^y & m^y
  \end{array}
 \right)
 \cdot \tauhat.
\ee
We can insert these data in the string network mass formula (\ref{eq:masssn}). 
It is clear that the first term in this expression is equal to the first term 
in (\ref{eq:masssquaredstfinal}). For the second term in (\ref{eq:masssn}) we 
find
\bea
 M^2_2 & = & L_x L_y \left( \begin{array}{c} 1 \\ 0 \\ 0 \\ 1 \end{array} 
 \right)^T \cM_{\tauhat} \otimes \cM_\gl \left( \begin{array}{c} 1 \\ 0 \\ 0 
 \\ 1 \end{array} \right) \ret
 & = & L_x L_y \left( \begin{array}{c} n^x \\ n^y \\ m^x \\ m^y \end{array} 
 \right)^T \cM_{\gl} \otimes \cM_\tau \left( \begin{array}{c} n^x \\ n^y \\ 
 m^x \\ m^y \end{array} \right)
 \label{eq:term1intermediate}
\eea
where in the second line we used $SL(2, \bZ)$-invariance and we interchanged the
two matrices $\cM$. Note that the fundamental torus has modular parameter
\be
 \tau = (c +i) \frac{L_y}{L_x},
\ee
and inserting this in $\cM_\tau$ we find
\be
 M_2^2 = \frac{1}{\gl_2} \left(
  \begin{array}{c}
   n^x L_x \\ n^y L_y \\ m^x L_x \\ m^y L_y
  \end{array}
 \right)^T
 \left(
  \begin{array}{cc}
   \Ghat       & \gl_1 \Ghat   \\
   \gl_1 \Ghat & |\gl^2| \Ghat
  \end{array}
 \right)
 \left(
  \begin{array}{c}
   n^x L_x \\ n^y L_y \\ m^x L_x \\ m^y L_y
  \end{array}
 \right),
\ee
which is precisely the second term in (\ref{eq:masssquaredstfinal}).

{\em The case with arbitrary determinant.}

\begin{figure}
 \begin{center}
  \epsfysize=4 cm
  \epsfbox{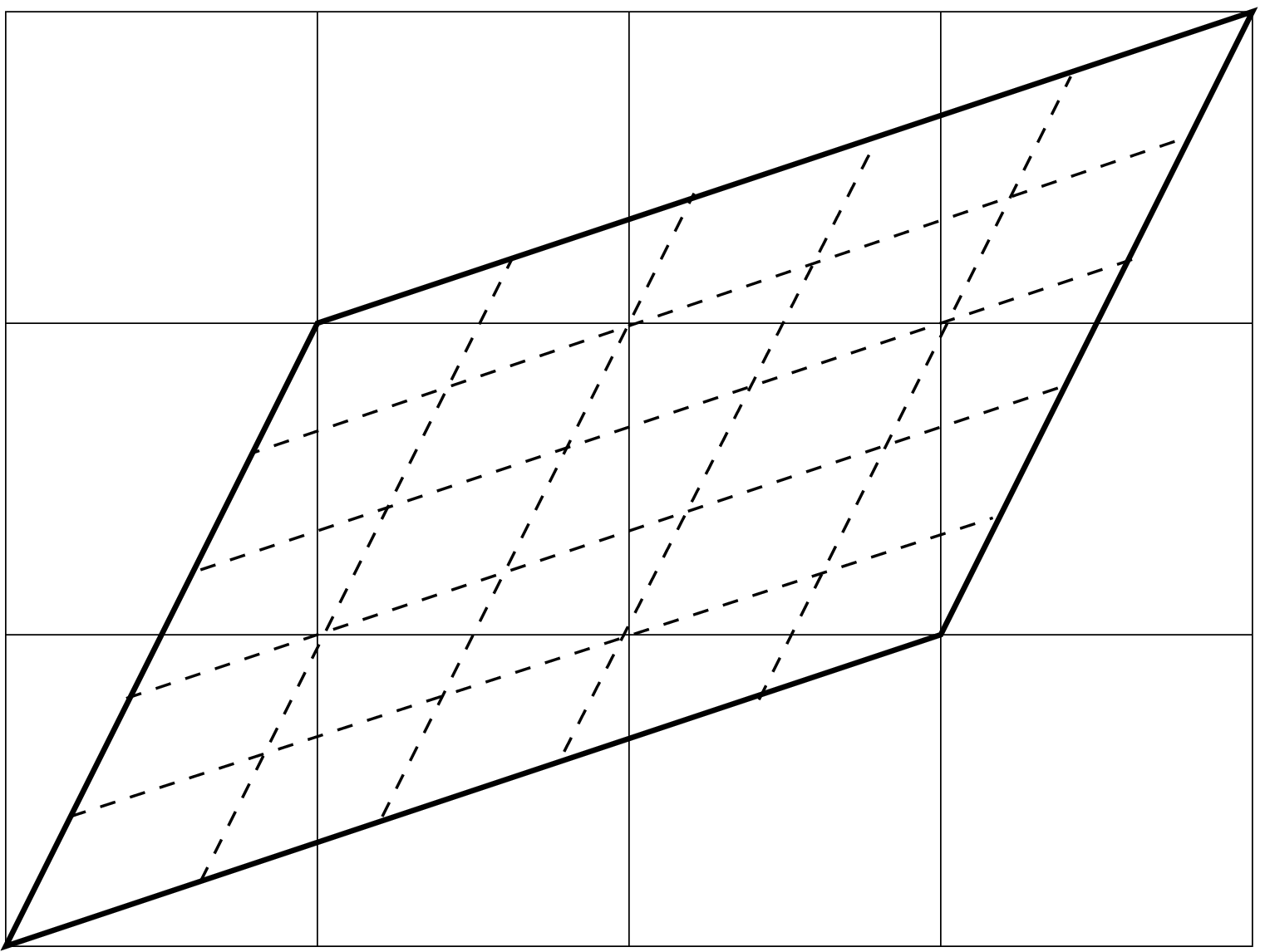}
 \end{center}
 \caption{Strings of winding numbers (3,1) and (1,2) span a parallelogram of
 area 5. The dotted lines are copies of the strings.}
 \label{fig:largetorus}
\end{figure}

Now suppose the determinant in equation (\ref{eq:determinant}) is $\gD \neq 1$.
Note that the determinant cannot be zero since in this case the $B$- and
$D$-fields would point in the same direction and there would be no angular
momentum on the supersheet. When we draw the F- and D-strings (again, before the
extra string grows) in the complex plane which covers the torus $T^2$, they span 
a parallelogram with an area which is $\gD$ times the original area of the 
torus; see figure \ref{fig:largetorus}. However, by periodicity this 
parallelogram contains $\gD$ parallel ``string bits'' of each type, at equal 
distances.

To obtain the mass squared of this network, we will calculate the mass squared
of one of the smaller parallelograms\footnote{Note that such a parallelogram 
is not actually a torus. However, since the mass of a string only depends on 
its length and type, we may just as well use the mass formula for string 
networks. Accordingly, we will speak of the ``modular parameter of a 
parallelogram'' even when it is not a torus.} in figure \ref{fig:largetorus} 
and multiply it by $\gD^2$ since $\gD$ of these parallelograms together form a 
copy of the original torus $T$.

Note that it is still true that the modular parameter of one of the small
parallelograms is
\be
 \tauhat = \frac{m^y \tau + m^x}{n^y \tau + n^x}.
 \label{eq:SL2Rtransform}
\ee
The problem is that in this case this is not an $SL(2, \bZ)$-transformation. 
However, to show the invariance of the mass formula (\ref{eq:masssn}), one does 
not need the integrality of the coefficients of the transformation, so the 
expression is actually $SL(2, \bR)$-invariant. Since we can multiply the 
coefficients in (\ref{eq:SL2Rtransform}) by an arbitrary constant this is 
enough, and we can write
\be
 \tauhat =
 \left\{
 \frac{1}{\sqrt{\gD}}
 \left(
  \begin{array}{rr}
   m^y & m^x \\
   n^y & n^x
  \end{array}
 \right)
 \right\}
 \cdot \tau
\ee
Using this transformation, we can carry out the same calculation as in the
previous paragraph, where now after the transformation the four-vectors obtain
an extra factor of $1/\sqrt{\gD}$. This gives an overall factor of $1/\gD$.
There is another overall factor of $1/\gD$ coming from the smaller area $A$ of
the torus. Putting all of this together, we see that these factors are exactly
canceled by the factor of $\gD^2$ which comes from the fact that we have to
calculate the mass squared of $\gD$ of these small tori. Hence we showed that we
again obtain the expression (\ref{eq:masssquaredstfinal}), and so also in this 
case the mass of the string network equals the mass of the supersheet.

\section{From string networks back to supersheets}
In the previous section, the string networks we considered consisted only of
$(1,0)$-, $(0,1)$- and $(1,1)$-strings. Of course, one would expect that other
string networks also blow up into supersheets. In fact, for a general
three-string network on a torus this is not hard to see. Starting from such a
network, we can always do an $SL(2, \bZ)$-transformation such that it consists 
of $(p,0)$-, $(0,q)$- and $(p,q)$-strings\footnote{Again, by a $(p,0)$-string,
we really mean $p$ $(1,0)$-strings.}. By exactly the same calculation as
before, we see that this network is equivalent to a supersheet with $p$ units of
electric flux along one direction and $q$ units of magnetic flux along the
other.

For a completely general string network on a torus, i.\ e.\ one containing many
string junctions, it is not easy to give an exact calculation showing the
equivalence to a specific supersheet. However, it is clear that we can always
find an effective $D$- and $B$-field inside such a network, where the direction
of the $D$-field is the direction of the fundamental strings, and the direction
of the $B$-field is that of the D-strings. Moreover, the flux of such a field
can be obtained by calculating the total F- or D-string charge of the strings
that are pointing through one of the cycles of the torus. (It is not hard to 
see that this is a constant number, independent of where we put this 
``measuring cycle''.)

Of course, we could set up a string network where all of the string junctions
are located in a small region of the torus. However, from entropy arguments one
would expect the string junctions to spread more or less equally along the
surface of the torus, thus making a local tension tensor and local
electromagnetic fields well-defined and more or less constant. This
``macroscopic state'' would then correspond to the supersheet with the correct
electric and magnetic fluxes.

\section{Conclusion}
In this paper we argued that the string networks in type IIB string theory blow 
up into cylindrical D3-branes which are the three-dimensional generalization of 
the supertubes of Mateos and Townsend. We argued that there is a many-to-one 
relation between string networks and supersheets, where only the effective 
electric and magnetic fluxes in the string network determine the supersheet we 
end up with.

It would be interesting to make the entropy arguments in the last section of
this paper, showing the equivalence of an arbitrary ``macroscopic'' string 
network to a supersheet, more precise. Another closely related and interesting 
question is how exactly the excitations of the single strings combine into 
``collective modes'' which correspond to the excitations of the D3-brane.

{\bf Acknowledgements.} We would like to thank Robbert Dijkgraaf for useful
discussions. EV is supported by DOE grant DE-FG02-91ER40571. MV is supported 
by the FOM/NWO programme ``Mathematical Physics''.


\begin{thebibliography}{99}
\bibitem{Mateos:2001st}
D.~Mateos and P.~K.~Townsend,
``Supertubes,''
Phys. Rev. Lett. 87 (2001) 011602
[hep-th/0103030].

\bibitem{Cho:2001sd}
J.-H.~Cho, P.~Oh,
``Super D-helix,''
Phys. Rev. D64 (2001) 106010
[hep-th/0105095].

\bibitem{Bak:2001js}
D.~Bak and S.-W.~Kim,
``Junctions of supersymmetric tubes,''
Nucl. Phys. B622 (2002) 95-114
[hep-th/0108207].

\bibitem{Bak:2002sb}
D.~Bak N.~Ohta and M.~M.~Sheikh-Jabbari,
``Supersymmetric brane-antibrane systems: matrix model description, stability
and decoupling limits,''
JHEP 0209 (2002) 048
[hep-th/0205265].

\bibitem{Mateos:2002ts}
D.~Mateos, S.~Ng and P.~K.~Townsend,
``Tachyons, supertubes and brane/anti-brane systems,''
JHEP 0203 (2002) 016
[hep-th/0112054].

\bibitem{Mateos:2002sc}
D.~Mateos, S.~Ng and P.~K.~Townsend,
``Supercurves,''
Phys. Lett. B583 (2002) 366-374
[hep-th/0204062].

\bibitem{Kruczenski:2002as}
M.~Kruczenski, R.~C.~Myers, A.~W.~Peet and D.~J.~Winters,
``Aspects of supertubes,''
JHEP 0205 (2002) 017
[hep-th/0204103].

\bibitem{Tamaryan:2002td}
S.~Tamaryan, D.~K.~Park and H.~J.~W.~M\"uller-Kirsten,
``Tubular D3-branes and their dualities,''
[hep-th/0209239].

\bibitem{Park:2003ss}
D.~K.~Park, S.~Tamaryan and H.~J.~W.~M\"uller-Kirsten,
``Supersphere,''
Phys. Lett. B551 (2003) 187-192
[hep-th/0210306].

\bibitem{Grandi:2002ss}
N.~E.~Grandi and A.~R.~Lugo,
``Supertubes and special holonomy,''
[hep-th/0212159].

\bibitem{Aharony:1998wb}
O.~Aharony, A.~Hanany and B.~Kol,
``Webs of (p,q) 5-branes, five dimensional field theories and grid diagrams,''
JHEP 9801 (1998) 002
[hep-th/9710116].

\bibitem{Sen:1998sn}
A.~Sen,
``String network,''
JHEP 9803 (1998) 005
[hep-th/9711130].

\bibitem{Kumar:2001bs}
A.~Kumar, R.~R.~Nayak and K.~L.~Panigrahi,
``Bound states of string networks and D-branes,''
Phys. Rev. Lett. 88 (2002) 121601
[hep-th/0108174].

\bibitem{Witten:1996bs}
E.~Witten,
``Bound states of strings and $p$-branes,''
Nucl. Phys. B460 (1996) 335-350
[hep-th/9510135].

\bibitem{Schwarz:1995ms}
J.~Schwarz,
``An $SL(2, \bZ)$ multiplet of type IIB superstrings,''
Phys. Lett. B360 (1995) 13-18; Phys. Lett. B364 (1995) 252
[hep-th/9508143].

\bibitem{Hofman:1998ud}
C.~Hofman, E.~Verlinde and G.~Zwart,
``U-Duality invariance of the four-dimensional Born-Infeld theory,''
JHEP 9810 (1998) 020
[hep-th/9808128].

\bibitem{Verlinde:1995ga}
E.~Verlinde,
``Global aspects of electric-magnetic duality,''
Nucl. Phys. B455 (1995) 211-228
[hep-th/9506011].

\end{thebibliography}
\end{document}